\documentclass{rspublic}
\usepackage{amsmath}
\usepackage{amssymb}
\usepackage{amsthm}
\usepackage{amsfonts}
\usepackage{epsfig} 
\usepackage{graphicx}
\usepackage{dcolumn}
\usepackage{bm}
\usepackage{ulem} 
\usepackage[]{natbib}
\usepackage[usenames]{color}

\def\supit#1{\raisebox{0.8ex}[8pt]{\footnotesize #1}\hspace{0.05em}}

\begin{document}

\title[Optimal path to extinction]{Converging towards the optimal path to extinction}

\author[I.B. Schwartz, E. Forgoston, S. Bianco, and L.B. Shaw]{Ira
  B. Schwartz\supit{1}, Eric Forgoston\supit{2}, Simone
  Bianco\supit{3}, \\and Leah B. Shaw\supit{4}}

\affiliation{\supit{1} Nonlinear Systems Dynamics Section,
       Plasma Physics Division,
       U.S. Naval Research Laboratory,
       Code 6792,
       Washington, DC 20375, USA\\
\supit{2} Department of
     Mathematical Sciences, 1 Normal Avenue, Montclair State University, Montclair, NJ 07043, USA
\\
\supit{3} Department of Bioengineering and Therapeutic Sciences, University of
California San Francisco, 1700 4th Street, San Francisco, CA 94158-2330, USA\\
\supit{4} Department of Applied Science,
       The College of William \& Mary,\\
       P.O. Box 8795,
       Williamsburg, VA 23187-8795, USA}

\label{firstpage}

\maketitle

\begin{abstract}{Extinction, Optimal Path, Finite-Time Lyapunov Exponents}
Extinction appears ubiquitously in many fields, including chemical reactions,
population biology, evolution, and epidemiology.  Even though extinction as a
random process is a rare event, its occurrence is
observed in large finite populations.  Extinction occurs when fluctuations due to random transitions act as
an effective force which drives one or more components or species to
vanish.  Although there are many random paths to an extinct state,
there is  an optimal path that maximizes the probability to
extinction.  In this article, we show that the optimal path
 is associated with the dynamical systems idea of having
maximum sensitive dependence to initial conditions.  Using the equivalence between the sensitive
dependence and the path to extinction, we show that the dynamical systems picture
of extinction evolves naturally toward the optimal path in several stochastic
models of epidemics.
\end{abstract}

\section{Introduction}
Determining the conditions for epidemic extinction is an important public
health problem.  Global eradication of an infectious disease has
rarely been achieved, but it continues to be a public health goal for
polio~\citep{WHO09} and many other diseases, including childhood diseases.  More commonly, disease
extinction, or fade out, is local and may be followed by a reintroduction of
the disease from
other regions~\citep{Anderson91,Grassly2005}.  Extinction may also occur for individual
strains of a multistrain disease~\citep{Minayev2009}, such as influenza or
dengue fever.  Since extinction occurs in finite populations, it  depends critically on local community
size~\citep{Bartlett1957,Bartlett1960,Keeling1997,Conlan2007}.
Moreover, it is important to know how parameters affect the chance of extinction for predicting the dynamics of outbreaks and
for developing control strategies to promote epidemic
extinction~\citep{dyscla08}.  The determination of extinction risk is
also of interest in related fields, such as evolution and ecology.  For example, in the neutral theory of ecology,
bio-diversity arises from the interplay between the introduction and
extinction of species~\citep{BanMar09,Azaele2006}.

In general, extinction occurs in discrete, finite populations undergoing
stochastic effects due to random transitions or perturbations.  The origins of stochasticity may be internal to
the system  or may arise from the external environment.  Small population size, low contact
frequency for frequency-dependent transmission, competition for
resources, and evolutionary pressure~\citep{DeCastro2005}, as well as heterogeneity in
populations and transmission~\citep{Lloyd2007}, may all be determining factors for
extinction to occur.

Extinction risk is affected by the nature and strength of the noise~\citep{Melbourne2008}, as well as other factors,
including outbreak amplitude~\citep{Alonso2006} and seasonal phase
occurrence~\citep{stolhu07}.  For large populations, the intensity of
internal population noise is generally small.  However, a rare,
large fluctuation can occur with non-zero probability and the
system may be able to reach the extinct state. For respiratory
diseases such as SARS, super-spreading may account
for these large stochastic fluctuations~\citep{Lloyd-Smith2005}.  Since the
extinct state is absorbing due to effective stochastic forces, eventual extinction is guaranteed when
there is no source of reintroduction~\citep{bartlett49,Allen2000,gar03}.  However,
because fade outs are usually rare events in large populations, typical time scales for extinction may be
extremely long.

Stochastic population models of finite populations, which include extinction processes, are
effectively described using the master
equation formalism.  Stochastic master equations are commonly used in
statistical physics when dealing with chemical reaction
processes~\citep{Kubo63} and predict probabilities of rare events occurring at certain
times. For many problems involving extinction in large populations, if the
probability distribution of the population is approximately stationary, the probability of extinction
is a function that decreases exponentially with increasing population size.  The exponent in this function scales as a
deterministic quantity called the action~\citep{Kubo1973}.  It can be shown that a
trajectory that brings the system to extinction is
very likely to lie
along a most probable path, called the optimal extinction trajectory
or optimal path.  It is a remarkable property that a
deterministic quantity such as the action can predict the probability of
extinction, which is inherently a stochastic process~\citep{sbdl09,dyscla08}.

Locating the optimal path is desirable because the quantity of interest, the
extinction rate, depends on the probability to traverse this path, and the
effect of a control strategy on extinction rate can be determined by its
effect on the optimal path~\citep{dyscla08}.  By employing an optimal path
formalism, we convert the stochastic problem to a mechanistic dynamical
systems problem.  In contrast to approaches based on diffusive processes
that are valid only in the limit of large system
sizes~\citep{Nasell1999,Lindholm2008}, this dynamical systems approach can give accurate
estimates for the extinction time even for small populations if
the action is sufficiently large.  Additionally, unlike other methods that are
used to estimate lifetimes, this approach enables one both to estimate
lifetimes and to draw conclusions about the path taken to extinction.  This more detailed
understanding of how extinction occurs may lead to new stochastic control
strategies~\citep{dyscla08}.

In this article, we show that locating
the optimal extinction trajectory can be achieved naturally by evolving a
dynamical system that converges to the optimal path.  The method is based on computing finite-time
Lyapunov exponents (FTLE), which have previously been used to find coherent structures in fluid
flows~\citep{hall00,hall01,hall02,shlema05,leshma07,brawig09}.  The FTLE
provides a measure of how sensitively the system's
future behavior depends on its current state.  We
argue that the system displays maximum sensitivity near the optimal
extinction trajectory, which enables us to dynamically evolve toward the optimal escape trajectory using FTLE
calculations.  For several models of epidemics that
contain internal or external noise
sources, we illustrate the power of our method to locate optimal extinction
trajectories.  Although our examples are taken from infectious disease models, the
 approach is general and is applicable to any extinction process or escape process.

\section{Stochastic modeling in the large population limit}\label{sec:opt_path}
To introduce our idea of dynamically constructing an optimal path to extinction in
stochastic systems, we show its application to a
stochastic Susceptible-Infectious-Recovered (SIR) epidemiological
model.   Details of the SIR model can be found in Sec.~1 of the electronic online supplement.  Figure~\ref{fig:SIR}
shows the probability density of extinction prehistory in the SI plane.
 The probability density was numerically computed using  $20,000$ stochastic SIR
trajectories that ended in extinction. Trajectories are aligned at
their extinction point.  From the extinction point, the prehistory of each
  trajectory up to the  last outbreak of infection is
  considered. Small fluctuations in the infectious population
  are not considered in identifying the last outbreak. In this way, we restrict the analysis to
  the interval between the last large outbreak of infection and the
  extinction point. The resulting (S,I) pairs of susceptible
  and infectious individuals are then binned and
  plotted in the SI plane~\citep{Eilers2004}.  The resulting discrete density has been color coded so
  that the brighter regions correspond to higher density of trajectories.
The figure shows that, among all the
paths that the stochastic system can take to reach the extinct state,
there is one path that has the highest probability of occurring. This is
the optimal path to extinction.   One can see that the optimal path to
  extinction lies on the peak of the probability density of the extinction prehistory. It should be noted that extinction for
  the stochastic SIR model
  has been studied previously~\citep{kammee08}.  

The optimal path can be obtained  using methods of
statistical physics. In figure~\ref{fig:SIR}, the numerical prediction
of the  entire optimal trajectory for the stochastic SIR system has
been overlaid on the probability density of extinction
  prehistory that was found using stochastic simulation. The
trajectory spirals away from the endemic state, with larger and
larger oscillations until it hits the extinct state. The agreement
between the stochastically simulated optimal path to
extinction and the predicted optimal path
is excellent.

The curve of figure~\ref{fig:SIR} is obtained by recasting
the stochastic problem in a deterministic form. The evolution of the
probability of finding a stochastic system in a given state $\mathbf{X}$
 at a given
time $t$ is described by the master equation~\citep{vanKampen_book}. Solving the master equation
would provide a complete description of the time evolution of the
stochastic system, but in
general it is a difficult task to obtain explicit solutions for the
master equation. Thus, one generally resorts to approximations to the
 solution; i.e., one considers an ansatz for the probability density. In this
case, since extinction of finite populations is a rare event, we will
be interested in examining events that occur in the tail of the probability
distribution.  Therefore, the distribution is assumed to take the form 
\begin{equation}\label{e:Action}
\rho(\mathbf{X},t) \approx \exp{(-N{\cal S}(\bm{x}))},
\end{equation}
where $\rho(\mathbf{X},t)$ is the probability density  of finding the system in
  the state $\mathbf{X}$ at time $t$,  $N$ is the size of the
population,  $\mathbf{x} = \mathbf{X}/N$ is the normalized state (e.g., in an
epidemic model, the fraction of the population in the various compartments), and $\cal{S}$ is a deterministic state function known in classical physics as
the action.  Equation~(\ref{e:Action}) describes the relationship between the action and the
probability density and is based on an assumption for how the probability scales with the population size. The action is the negative of the  natural log of the stationary probability distribution
divided by the population size.  Therefore, the probability (if we normalize the
population) is roughly given by the exponential of the action.
  Intuitively, equation~(\ref{e:Action}) expresses the assumption that the
  probability of occurrence of extreme events, such as extinction, lies in the tails
of the probability distribution, which falls steeply away from the
steady state.

This approximation leads to a conserved quantity that is
  called the Hamiltonian~\citep{gang87}.  From the Hamiltonian, one can find a set of conservative ordinary differential equations
(ODEs) that are known as Hamilton's equations.   These ODEs describe the
time evolution of the system in terms of state variables $\mathbf{x}$,
which are associated with the population in each compartment.  For the SIR example, $\mathbf{x}$ is the
vector $\left <S, I, R \right >$.  In addition to the state variables, the
equations contain conjugate momenta
variables, $\mathbf{p_x}$.  The conjugate momenta, or noise, account for the uncertainty
associated with the system being in a given state at a given time due to
the stochastic interactions among the individuals of the population. These ODEs can be
constructed from information in the master equation about the possible
transitions and transition rates in the system.  Details can be found in~\ref{sec:app-tlf}.

Integration of the ODEs with the
appropriate boundary conditions will then give the
optimal evolution of the system under the influence of the noise.  Boundary
conditions are chosen to be fixed points of the system.  A typical case is
shown schematically in figure~\ref{fig:hyperbolic}a.  Deterministically, the
endemic state is attracting and the extinct state repelling.  However,
introducing stochasticity allows the system to leave the deterministic
manifold along an unstable direction of the endemic state, corresponding to
nonzero noise.  Stochasticity leads to an additional extinct state which
  arises due to the general non-Gaussian nature of the noise.   For the
extinction process of figure~\ref{fig:SIR}, boundary conditions were
the system leaving the endemic steady state and asymptotically
approaching the stochastic extinct state.

In general, the optimal extinction path is an unstable dynamical object, and
this reflects extinction as a rare event. This has led many authors
to consider how extinction rates scale with respect to a parameter
close to a bifurcation
point~\citep{Doering2005,kammee08,Kamenev2008b,dyscla08}, where the
dynamics are very slow. For an
epidemic model this means that the reproductive rate $R_0$ should be
greater than but very close to $1$. However, most real diseases have $R_0$
larger than 1.5, which translates into a faster growth rate from the
extinct state. In general, in order to obtain analytic scaling
results, one must obtain the ODEs for
the optimal path either analytically (using the
classical theory of large fluctuations  mentioned within this section and
described  in detail in~\ref{sec:app-tlf}) or
numerically (using shooting methods for boundary value problems). This
task may be impossible or extremely cumbersome, especially when the
system is far from the bifurcation point. In the following section we
demonstrate how to evolve naturally to the optimal path to extinction
using a dynamical systems approach.

\section{Finite-time Lyapunov exponents }\label{sec:FTLE}
We consider a velocity field which is defined over a finite  time
interval and is given by Hamilton's equations of motion.  Such a velocity field, when starting from an initial condition,
has a unique solution.  The continuous dynamical system has quantities, known as Lyapunov
exponents, which are associated with the trajectory of the system
in an infinite time limit, and which measure the average growth
rates of the linearized dynamics about the trajectory.  To find the
finite-time Lyapunov exponents (FTLE), one computes the Lyapunov exponents
on a restricted finite time interval.  For each initial condition, the
exponents provide a measure of its sensitivity to small perturbations.
Therefore, the FTLE is
a measure of the local sensitivity to initial data.  Details regarding the
FTLE can be found in Sec.~2 of the electronic online supplement.

The  FTLE measurements can be shown to exhibit ``ridges'' of local maxima.  The ridges
of the field indicate the location of attracting (backward time FTLE
field) and repelling (forward time FTLE field)
structures.  In two-dimensional (2D) space, the ridge is a curve which locally maximizes
the FTLE field so that transverse to the ridge one finds the FTLE to
be a local maximum.  What is remarkable is that the FTLE ridges correspond to
the optimal path trajectories, which we heuristically argue in Sec.~3 of the
electronic online supplement.  The basic idea is that since the optimal path is inherently
unstable, the FTLE shows that, locally, the path is also the most sensitive
to initial data.  Figure~\ref{fig:hyperbolic}b shows a schematic that
demonstrates why the optimal path has a local maximum to sensitivity.  If one chooses
an initial point on either side of the path near the endemic state, the two trajectories
will separate exponentially in time.  This is due to the fact that both extinct
and endemic states are unstable, and the connecting trajectory defining the
path is unstable as well.  Any initial points starting near the optimal path
will leave the neighborhood in short time.

\section{Evolving towards  the optimal path using FTLE}\label{sec:apps}

We now apply our theory of dynamical sensitivity to the problem of locating
optimal paths to extinction for several examples.  We consider the case of
internal fluctuations, where the noise is not known a priori, as well as the
case of external noise.  In each case, the interaction of
the noise and state of the system begins   by finding the equations of motion that describe the
unstable flow. These
equations of motion are then used to compute the ridges corresponding to
maximum FTLE, which
in turn correspond to the optimal extinction paths~\citep{fbss10pre}.

\subsection{Extinction in a  Branching - Annihilation Process}\label{sec:birth_death}

For an example of a system with internal fluctuations which has an
analytical solution, consider extinction in the stochastic
branching-annihilation process
\begin{equation}
A\xrightarrow{\lambda}2A\quad{\rm
  and}\quad2A\xrightarrow{\mu}\emptyset,\label{e:bd}
\end{equation}
where $\lambda$ and $\mu>0$ are constant reaction rates~\citep{elgkam04,askame08}.
Equation~(\ref{e:bd}) is a single species birth-death process and
can be thought of as a simplified form of the Verhulst logistic model
for population growth~\citep{nase01}.  The mean field equation for the
  average number of individuals $n$ in the infinite population limit is given by $\dot{n}=\lambda n -\mu n^2$. The stochastic process given by
equation~(\ref{e:bd}) contains intrinsic noise which arises from the randomness of the reactions and the fact
that the population consists of discrete individuals.  This intrinsic
noise can generate a rare sequence of events that causes the
system to evolve to the extinct state. The probability $P_{n}(t)$ to observe,
at time $t$, $n$ individuals is governed by the master
equation \begin{equation}
\dot{P}_{n}=\frac{\mu}{2}\left[(n+2)(n+1)P_{n+2}-n(n-1)P_{n}\right]+\lambda\left[(n-1)P_{n-1}-nP_{n}\right].\label{e:bd_master}\end{equation}

The Hamiltonian associated with this system is
\begin{equation}
H(q,p)=\left(\lambda(1+p)-\frac{\mu}{2}(2+p)q\right)qp,\label{e:bd_Ham}
\end{equation}
where $q$ is a conjugate coordinate related to $n$ through a transformation~\citep{askame08},
and $p$ plays the role of the momentum.
 The equations of motion are given by
\begin{subequations}
\begin{flalign}
\dot{q}= & \frac{\partial H}{\partial p}=q[\lambda(1+2p)-\mu(1+p)q],\label{e:bd_qdot}\\
\dot{p}= & -\frac{\partial H}{\partial q}=p[\mu(2+p)q-\lambda(1+p)].\label{e:bd_pdot}\end{flalign}
 \end{subequations}

The mean field is retrieved in equation~(\ref{e:bd_qdot}) when $p=0$ (no fluctuations or noise).
  The Hamiltonian
has three zero-energy
curves.  The first is the mean-field zero-energy line $p=0$ (no fluctuations), which
contains two unstable points $h_{1}=(p,q)=(0,\lambda/\mu)$
and $h_{0}=(p,q)=(0,0)$.  The second is the extinction line $q=0$ (trivial solution),
which contains another unstable point $h_{2}=(p,q)=(-1,0)$.
The third zero-energy curve is non-trivial and is
\begin{equation}
q=\frac{2\lambda(1+p)}{\mu(2+p)}.\label{e:bd_op}
\end{equation}
The segment of the curve given by equation~(\ref{e:bd_op}) which lies
between $-1\leq p\leq0$ corresponds to a (heteroclinic) trajectory
which exits, at $t=-\infty$, the point $h_{1}$ along
its unstable manifold and enters, at $t=\infty$, the  point
$h_{2}$ along its stable manifold.  This trajectory is
the optimal path to extinction and describes the most probable
sequence of events which evolves the system from a quasi-stationary
state to extinction~\citep{askame08}.

To show that the FTLE evolves to the optimal path, we calculate the FTLE field using
the system of Hamilton's equations given by equations~(\ref{e:bd_qdot})-(\ref{e:bd_pdot}).  Figure~\ref{fig:FTLE_bd_SIS}a
shows both the forward and backward FTLE plot computed
for the finite time $T=6$, with $\lambda=2.0$ and $\mu=0.5$.  In this
  example, as well as the following two examples, $T$ was chosen to be sufficiently
  large so that one obtains a measurable exponential separation of trajectories. In figure~\ref{fig:FTLE_bd_SIS}a,
one can see that the optimal path to extinction is given by the ridge
associated with the maximum FTLE.  In fact, by overlaying the forward and backward FTLE fields, one can see all three zero-energy curves including the optimal
path to extinction.
Also shown in figure~\ref{fig:FTLE_bd_SIS}a
are the analytical solutions to the three zero-energy curves given
by $p=0$, $q=0$, and equation~(\ref{e:bd_op}).  There is excellent agreement
between the analytical solutions of all three curves and the ridges which
are found through numerical computation of the FTLE flow fields.

It is possible to compute analytically the action along the optimal path for a
range of $\lambda /\mu$ values.  Using equation~(\ref{e:bd_op}), it is easy to
show that the action $\cal{S}$ is
\begin{equation}
{\cal{S}}=2(1-\ln{2})\frac{\lambda}{\mu}.\label{e:bd_action}
\end{equation}
It is clear from equation~(\ref{e:bd_action}) that the action scales linearly
with $\lambda /\mu$.

\subsection{SIS Epidemic Model - External Fluctuations}\label{sec:SIS_L}
We now consider the well-known problem of extinction in a Susceptible-Infectious-Susceptible
(SIS) epidemiological model, which is a core model for the basis of many
recurrent epidemics.  The SIS model is described by the following system of
equations:
\begin{subequations}
\begin{flalign}
\dot{S}=&\mu-\mu S +\gamma I - \beta IS,\label{e:SIex_a}\\
\dot{I}=&-(\mu +\gamma)I +\beta IS,\label{e:SIex_b}
\end{flalign}
\end{subequations}
where $\mu$ denotes a constant birth and death rate, $\beta$ represents the
contact rate, and $\gamma$ denotes the recovery rate.  Assuming the total
population size is constant and can be normalized to $S+I=1$, then
equations~(\ref{e:SIex_a})-(\ref{e:SIex_b}) can be rewritten as the the following
one-dimensional (1D) process for the fraction of infectious
individuals in the population:
\begin{equation}
\dot{I}=-(\mu+\gamma)I+\beta
I(1-I).\label{e:1DSIS_det}
\end{equation}
The stochastic version of equation~(\ref{e:1DSIS_det}) is given as
\begin{equation}
\dot{I}=-(\mu+\gamma)I+\beta
I(1-I)+\sigma\eta(t)=F(I)+\sigma\eta(t),\label{e:1DSIS_stoch}
\end{equation}
where $\eta(t)$ is uncorrelated Gaussian noise with zero mean and
  $\sigma$ is the standard deviation of the noise intensity.  The noise models random migration to and from another
population~\citep{Alonso2006,Doering2005}.

Equation~(\ref{e:1DSIS_stoch})
has two equilibrium points given by $I=0$ (disease-free
state) and $I=1-(\mu+\gamma)/\beta$ (endemic
state).  Using the Euler-Lagrange equation of motion~\citep{Goldstein2001} from the
Lagrangian determined by equation~(\ref{e:1DSIS_stoch})
($L(I,\dot{I})=[\eta(t)]^{2}=[\dot{I}-F(I)]^{2}$) along with the
boundary conditions given by the extinct and endemic states,
one finds that the optimal path to extinction (as well as its counterpart
path from the disease-free state to the endemic state) is given by
$\dot{I}=\pm F(I)$.

\begin{figure}
 \begin{center}
 \begin{minipage}[c]{0.49\linewidth}
 \includegraphics[width=6.25cm]{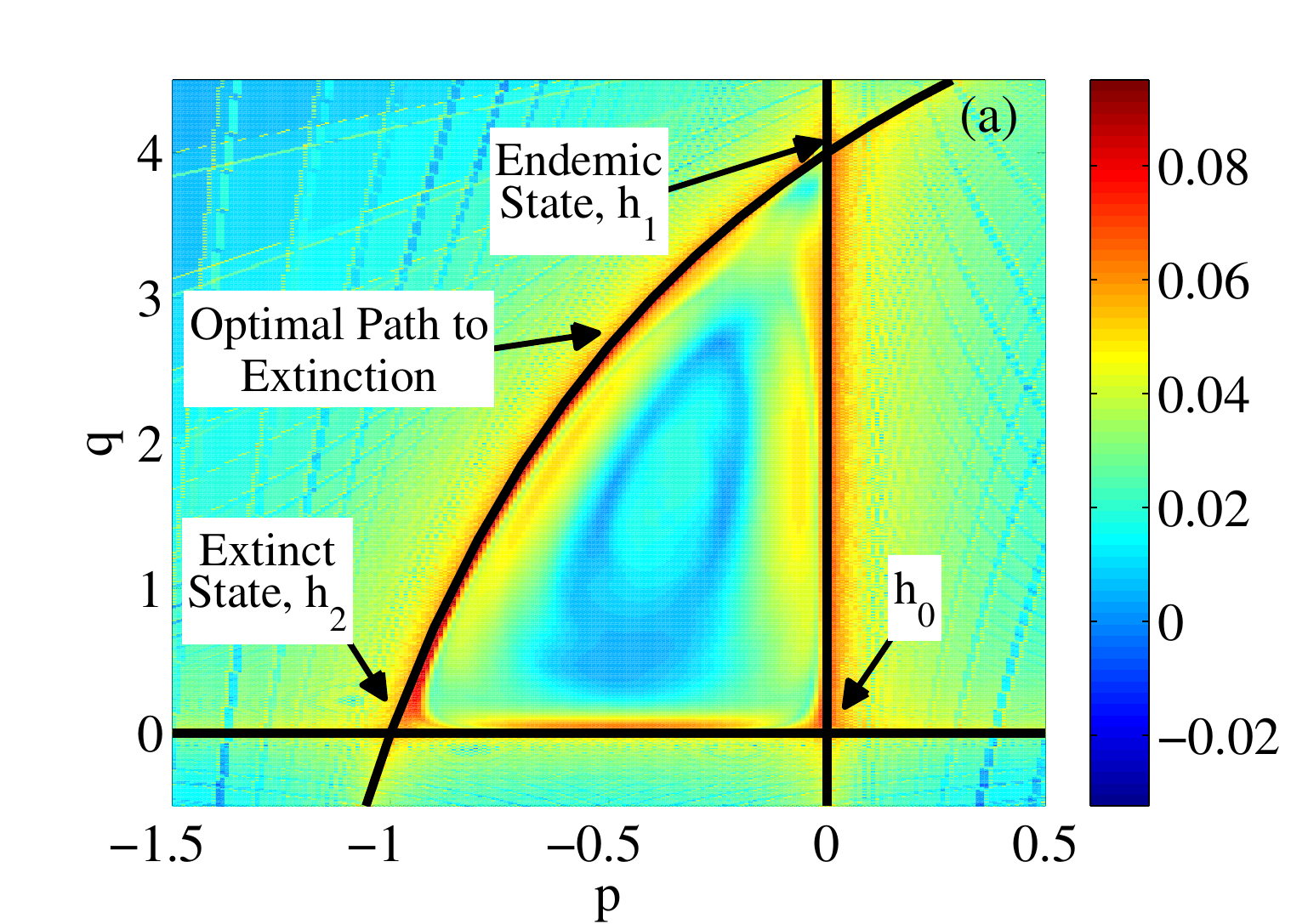}
 \end{minipage}
 \begin{minipage}[c]{0.49\linewidth}
 \includegraphics[width=6.25cm]{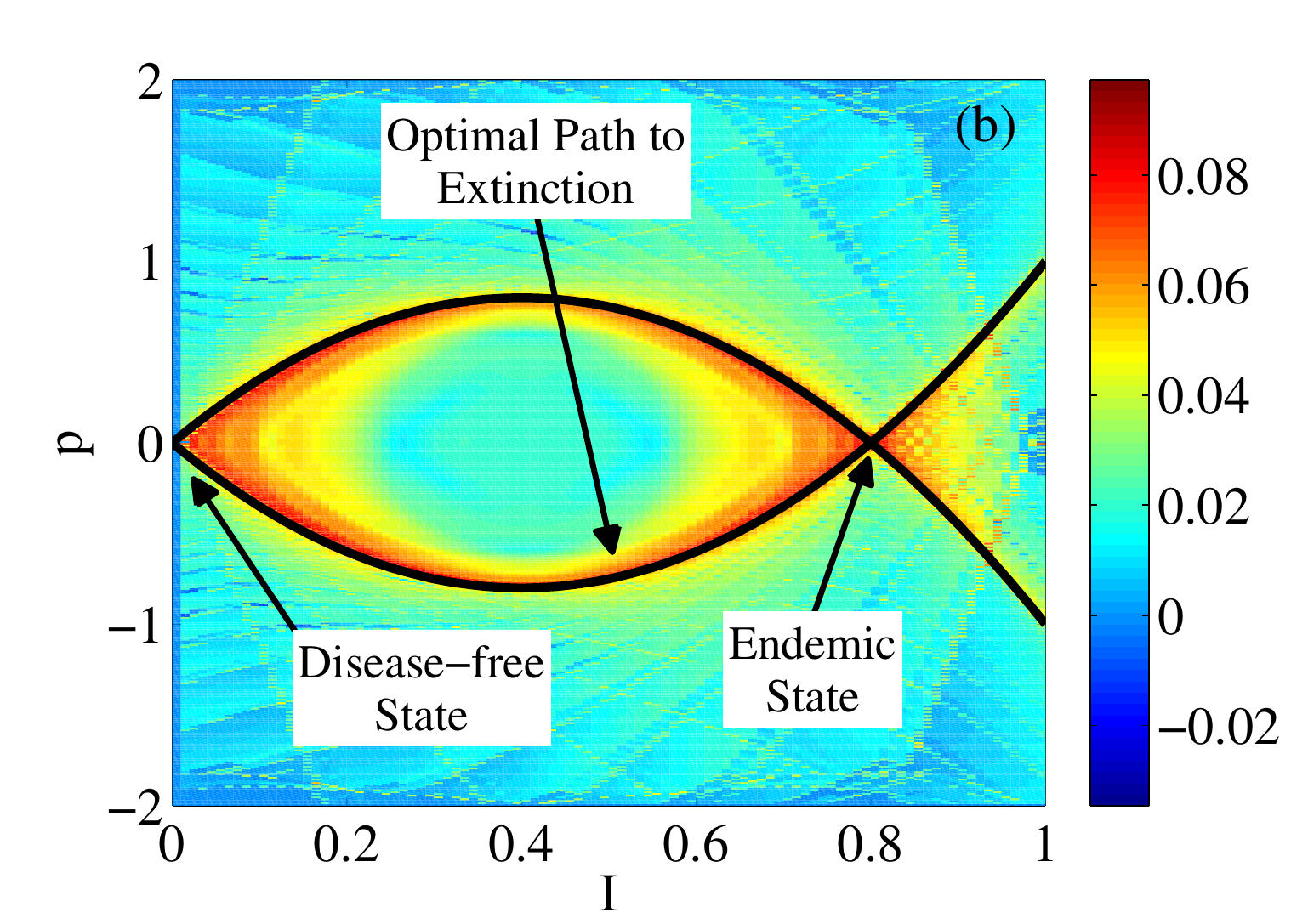}
 \end{minipage}
 \caption{ (a) Forward and backward FTLE flow fields for the branching-annihilation process computed using the
     Hamiltonian given by equation~(\ref{e:bd_Ham}) with $\lambda=2.0$ and
     $\mu=0.5$.  The integration time is $T=6$ with an integration step
     size of $t=0.1$ and a grid resolution of $0.01$ in both $q$ and
     $p$ (momentum).  The three zero-energy
     curves are given by the ridges of maximal FTLE and are overlaid with the analytical solution
     of these curves given by $p=0$, $q=0$, and equation~(\ref{e:bd_op}).  (b)
     Forward and backward FTLE flow fields for the SIS
     epidemic model with external fluctuations.  The flow fields were computed using
     the Lagrangian given by equation~(\ref{e:1DSIS_stoch}) with $\beta=5.0$
     and $\kappa=\mu +\gamma =1.0$.
     The integration time is $T=5$ with an integration step size of $t=0.1$
     and a grid resolution of $0.01$ in both $I$ and $p$ (momentum).
     The optimal path to extinction from the endemic state to the
       disease-free state
     and its counterpart optimal path from the disease-free state to
the endemic state (found by computing the backward FTLE field) are given by the ridges of maximal FTLE
     and are overlaid with
     the analytical solution of the optimal paths.}\label{fig:FTLE_bd_SIS}
 \end{center}
 \end{figure}

As in the first example, one can numerically compute the optimal path to extinction
using the FTLE.  Figure~\ref{fig:FTLE_bd_SIS}b
shows the forward and backward FTLE plot computed
for $T=5$, with $\beta=5.0$ and $\kappa=\mu +\gamma =1.0$. Note that we can
consider the combination $\mu+\gamma$ since the Lagrangian depends only on
the combination rather than on
either parameter separately.  In figure~\ref{fig:FTLE_bd_SIS}b,
one can see that the optimal path from the endemic state to the disease-free
state is given by the ridge associated with the local maximum FTLE.  Also
shown in figure~\ref{fig:FTLE_bd_SIS}b is the counterpart optimal path from the disease-free state to
the endemic state (found by computing the backward FTLE field).  In addition, the agreement with the analytical prediction is excellent, as shown in figure~\ref{fig:FTLE_bd_SIS}b.

If one solves equation~(\ref{e:1DSIS_stoch}) for $I(t)$ and substitutes both
$I(t)$ and $\dot{I}(t)$ into the Lagrangian, then one can analytically find an
expression for
the action along the optimal path.  The expression for the action is a function of $\kappa$ and
the reproductive number $R_0$ and is given by
\begin{equation}
{\cal{S}}=\frac{2\kappa (R_0-1)^3}{3R_0^2}, \label{e:SIS_L_action}
\end{equation}
where $R_0=\beta /\kappa$.

\subsection{SIS Epidemic Model - Internal Fluctuations}\label{sec:SIS_H}

\begin{figure}
 \begin{center}
 \includegraphics[width=8.5cm]{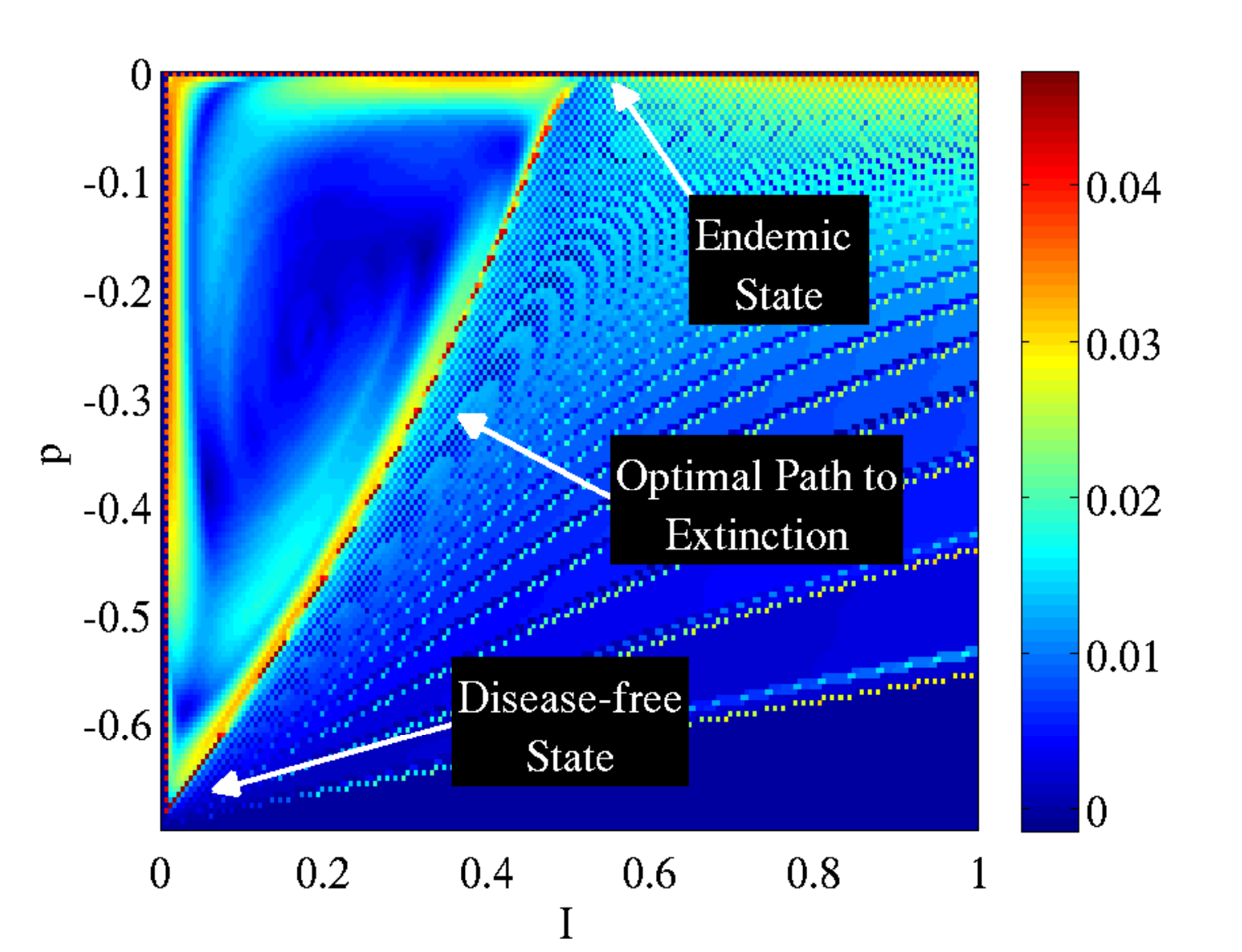}
   \caption{ FTLE flow field for the SIS epidemic model with internal fluctuations computed
     using the Hamiltonian given by equation~\ref{e:Ham_exp} with $\beta=2.0$ and
     $\kappa=\mu +\gamma =1.0$.  The integration time is $T=10$ with an integration
     step size of $t=0.1$ and a grid resolution of $0.005$ in both $I$
     and $p$ (momentum).  The optimal path to extinction is given by the ridge of maximal
     FTLE.}\label{fig:FTLE_Ham}
 \end{center}
 \end{figure}

 We next consider the 1D stochastic version of
the SIS epidemic model for a finite population with internal fluctuations using the transition rates
found in Sec.~4 of the electronic online supplement. 
Using the formalism of~\cite{gang87}, one then has the following
Hamiltonian associated with this model:
\begin{equation}
H(I,p)=(\mu+\gamma)I(e^{-p}-1)+\beta I(1-I)(e^{p}-1),\label{e:Ham_exp}
\end{equation}
where $I$ is the fraction of infectious individuals and $p$ is the momentum.
Internal fluctuations arise from the random interactions of the population.  Although there is no analytical solution for the optimal path to extinction,
we can once again
determine the optimal path by computing the FTLE flow field associated
with this system.  Figure~\ref{fig:FTLE_Ham} shows the forward FTLE
plot computed using Hamilton's equation of motion for
$T=10$, with $\beta=2.0$ and $\kappa=\mu + \gamma =1.0$, and as in previous examples, the optimal path to extinction from the endemic state to the disease-free
state is apparent.  Note that the non-zero momentum corresponding to the
extinct state qualitatively agrees with similar boundary conditions found in~\cite{dyscla08} and is associated with non-zero probability flux.

Once again, it is possible to compute the action along the optimal path for a
range of values of the reproductive number $R_0$.  In contrast to the prior
two examples, here the action must be computed numerically.  Moreover, even
the optimal path must be found numerically using the FTLE plot generated for
each value of $R_0$.

Given an $R_0$, we computed the forward FTLE flow field using a grid
resolution of $0.005$ in both position and momentum space.  To find the
optimal path corresponding to the ridge of maximal FTLE values, we let the
deterministic, endemic steady state be the starting point ${\bf z}_0$ of the
path.  Then a second point ${\bf z}_1$ on the path was found by locating the maximum of the FTLE values in
an arc of radius $20$ grid points spanning nearly $\pi$ radians for negative
momentum values and originating at ${\bf z}_0$.  Subsequent points
${\bf z}_{n+1}$ were found by taking the maximum FTLE value on an arc of radius
$10$ grid points spanning nearly $\pi$ radians originating at the most recent
point ${\bf z}_n$ and centered around the vector ${\bf z}_n-{\bf z}_{n-1}$ until
the extinction line was reached.  The complete optimal path was estimated from the
${\bf z}_n$
using cubic spline interpolation.  Once the optimal path had been found, the action was
computed by numerically integrating the Lagrangian along this path.

\begin{figure}
 \begin{center}
 \begin{minipage}[c]{0.49\linewidth}
 \includegraphics[width=6.25cm]{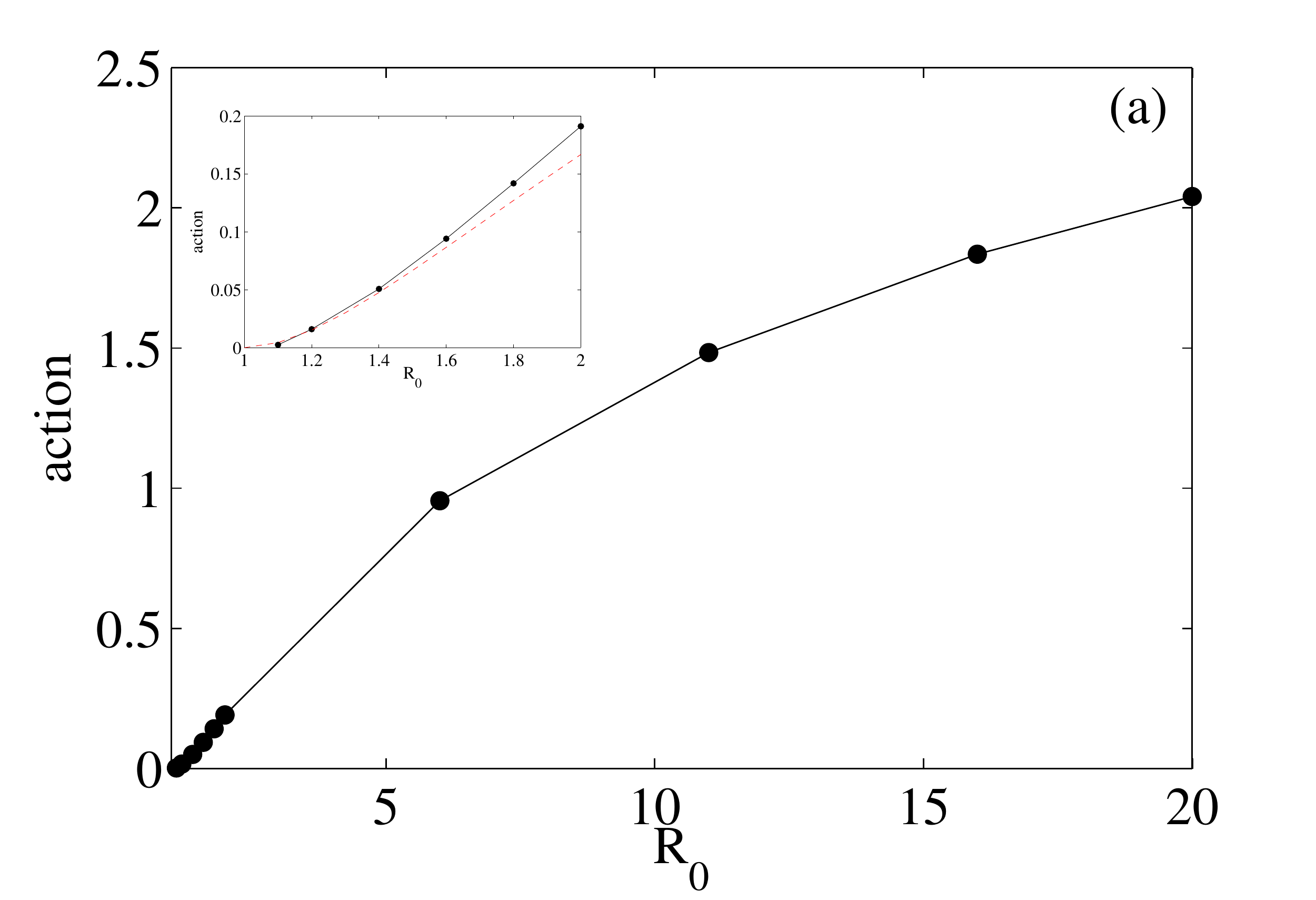}
 \end{minipage}
 \begin{minipage}[c]{0.49\linewidth}
 \includegraphics[width=6.25cm,height=4.2cm]{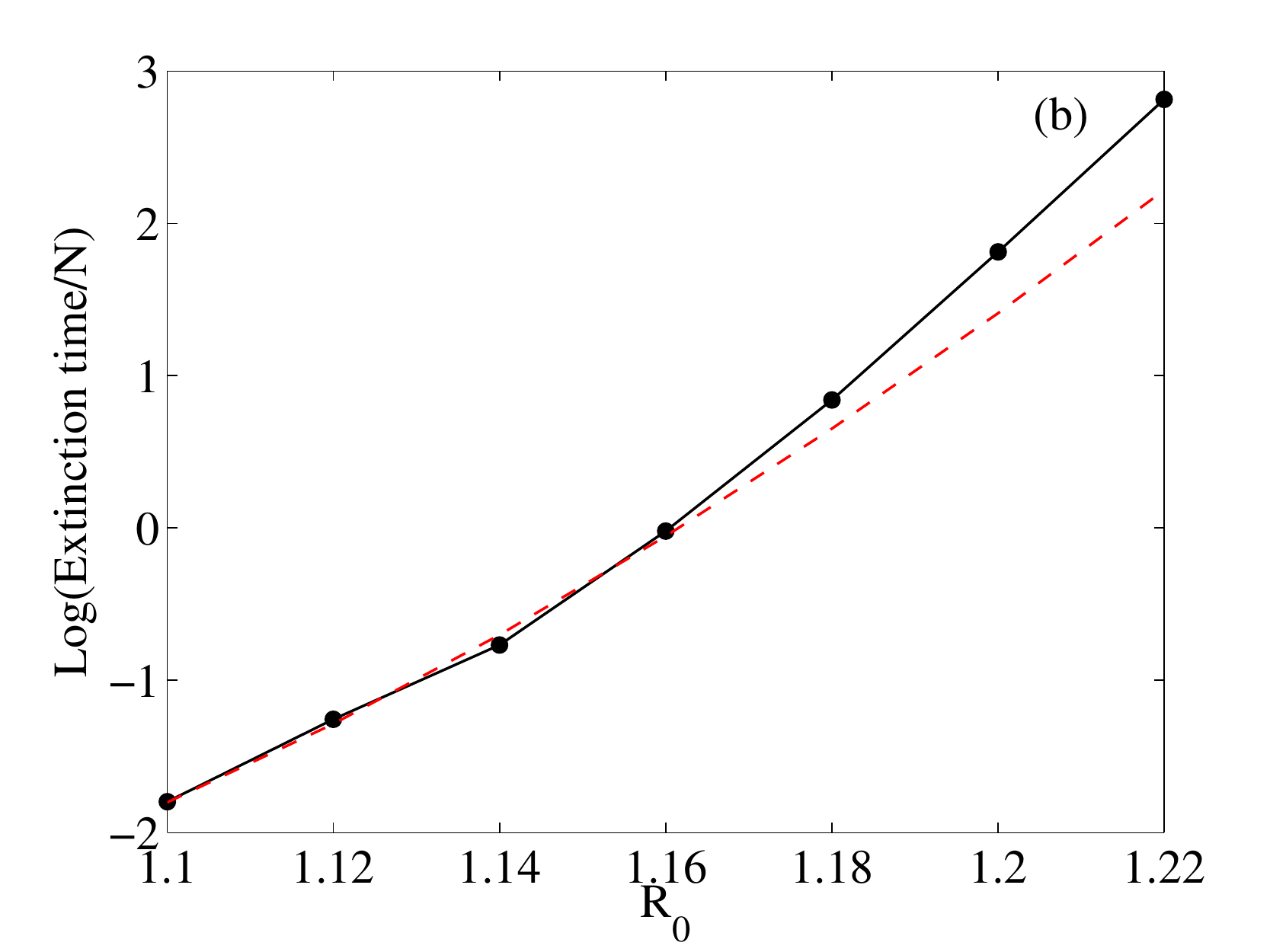}
 \end{minipage}
   \caption{\label{fig:action_v_R0} (a) Numerically computed action (integrated
     along the optimal path found numerically from the FTLE flow field) versus
     reproductive number $R_0$ for the SIS epidemic model with internal
     fluctuations.  The inset shows a portion of the graph near $R_0=1$.  The numerically computed action is given by the black
     points, while the dashed, red curve shows an asymptotic scaling result
     that is valid near $R_0=1$, and is given by
     $S(R_{0})=(R_0-1)^2/[R_0(1+R_0)]$ (see Sec.~5 of the electronic online
     supplement).  (b) Numerically simulated (solid curve with black points) mean
   extinction time versus reproductive number for the SIS epidemic model
   with internal noise and the analytical prediction (dashed, red curve) found
   using the asymptotic scaling law that is valid near $R_0=1$.}
 \end{center}
 \end{figure}

Figure~\ref{fig:action_v_R0}a shows the numerically computed action versus
reproductive number over the range $1.1 \leq R_0 \leq 20$.   The inset of
figure~\ref{fig:action_v_R0}a shows the portion of figure~\ref{fig:action_v_R0}a from $1.1 \leq R_0
\leq 2$.  Also shown in the inset of figure~\ref{fig:action_v_R0}a is an analytical, asymptotic scaling result
that is valid for values of $R_0$ close to unity.  As can be see in figure~\ref{fig:action_v_R0}a,
there is good agreement between the numerically computed action and the
analytically computed action near $R_0=1$.  Details of the derivation of the
analytical scaling law can be found in Sec.~5 of the electronic online
supplement.

 Figure~\ref{fig:action_v_R0}b shows the numerically simulated mean
  extinction time versus reproductive number for the 1D stochastic SIS model
  for a finite population (see Sec.~\ref{sec:apps}\ref{sec:SIS_H}).  Also
  shown in figure~\ref{fig:action_v_R0}b is the analytical prediction found
  using the previously mentioned scaling law derived for values of $R_0$ close
  to unity.  As one can see, the agreement is excellent.

\section{Conclusions}\label{sec:conc}
In all of the examples of Sec.~\ref{sec:apps}, we have shown that the optimal path to
extinction
is equated with having a (locally) maximal sensitivity to initial
conditions.  Even though there exist many possible paths to extinction, the
dynamical systems approach converges to the path that maximizes
extinction.  The parameter values chosen for the three examples are
such that the extinct and endemic states are far away from each other.
Therefore, in general, there will be no possible approximate analytical
treatment as performed in~\cite{dyscla08}.  In addition, we have shown how to
constructively compute numerically the action for a wide range of reproductive
numbers.  Our method allows for the computation of extinction times and can be
extended to high-dimensional problems.

Because the method is general, and unifies dynamical
systems theory with the probability of extinction, we expect that any
system found in other fields can be understood using this
approach.  Indeed, in problems of general extinction, it is now possible to
evolve naturally to the optimal path, and thus predict the path that
maximizes the probability to extinction.  Future work in this area
will include improved sampling methods to find the optimal path more
efficiently in higher dimensional models. Specific applications of optimal path
location in the future will include  spatio-temporal extinction
processes such as those that occur in pre-vaccine measles~\citep{Keeling1997,finkenstadt} and multi-strain extinction in diseases
such as dengue~\citep{cummings}.

\begin{figure}
 \begin{center}
 \includegraphics[width=8.5cm,height=5cm]{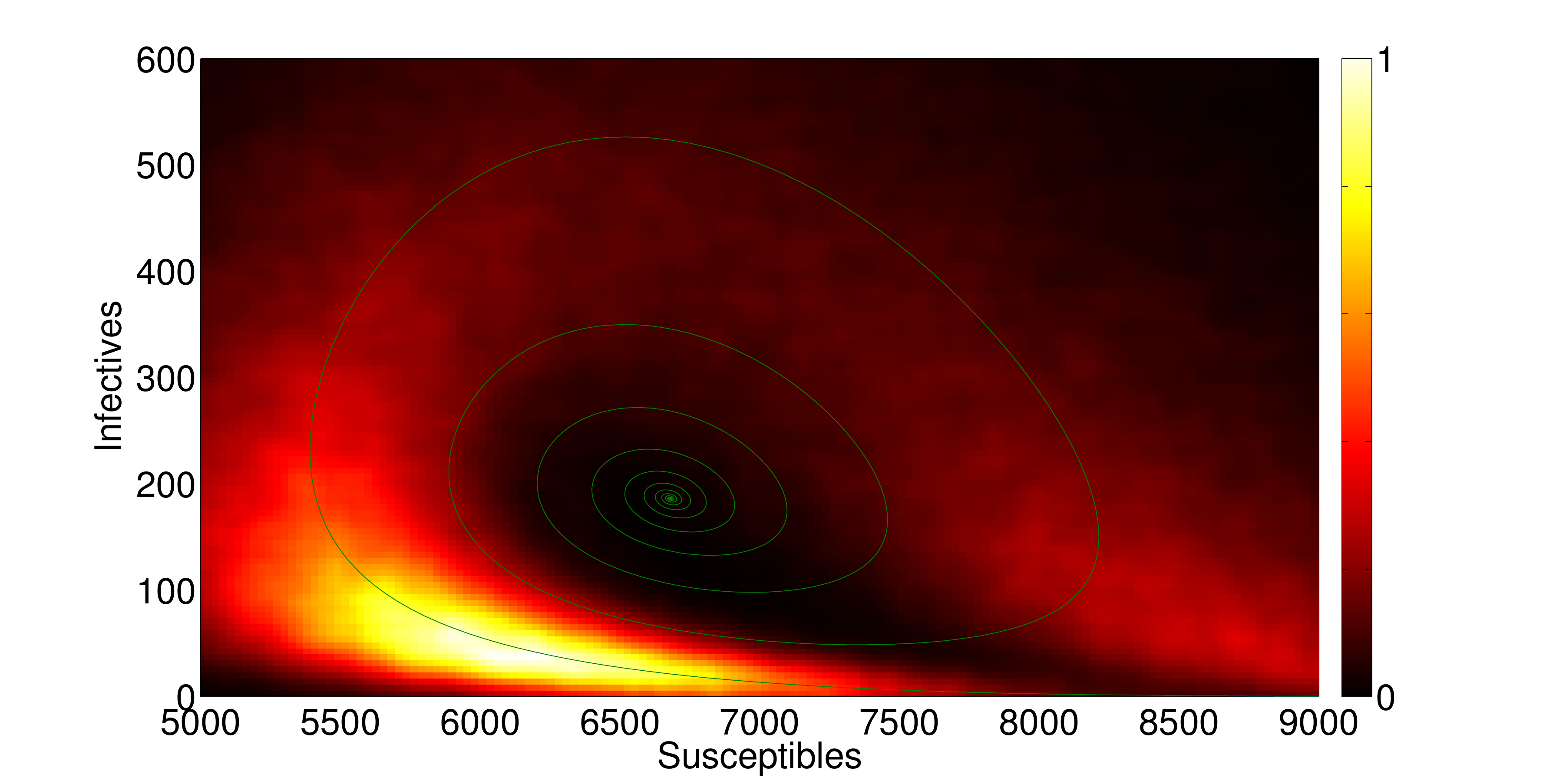}
 \caption{Probability density of extinction prehistory and
     the optimal path to extinction for the stochastic SIR
   epidemiological model. Colors indicate the probability density
   (the colorbar values have been normalized, with lighter colors corresponding to higher probability) for $20,000$
 stochastic realizations. The results were
   computed using Monte Carlo simulations, and details of the sampling of the
 trajectories are described in the text.  The curve is the numerically
   predicted optimal path to extinction.  Note that the optimal path to
  extinction lies on the peak of the probability density of extinction prehistory. The
 population is $3\times 10^6$ individuals, with $R_0 \approx 15$ (contact rate
 $\beta = 1500$, recovery rate $\gamma = 100$, birth-death rate $\mu =
 0.2$). }\label{fig:SIR}
 \end{center}
 \end{figure}

Finally, the optimal path method may lead to novel
  monitoring and control
strategies.  
In one biological
  application, knowledge of the most probable path to extinction may help with
  monitoring an environment and  with
  providing an estimate of the
  likelihood of extinction based on where the population lies 
    relative to the path.  It is known that once a trajectory has left the
  neighborhood of the endemic state, most paths to extinction occur near the
  optimal path.  This phenomenon can be seen in figure~\ref{fig:SIR}, where the
  optimal path lies on the peak of the probability density of extinction
  prehistory.  Therefore, the optimal path provides a good location to monitor
  the system for possible extinction events.  Furthermore,
  although the momenta (noise effects) are not directly observable, it may be
  possible to infer them dynamically~\citep{Luchinsky2008} from time series data
  of observable quantities  in conjunction with the
  equations for the time evolution of position and momentum variables.
  Knowledge of where a system lies in position-momentum space could provide an
  estimate for how quickly a desired epidemic extinction could occur or could
  provide the
  risk of extinction for a species one wishes to conserve.

\begin{figure}
 \begin{center}
 \includegraphics[width=6cm]{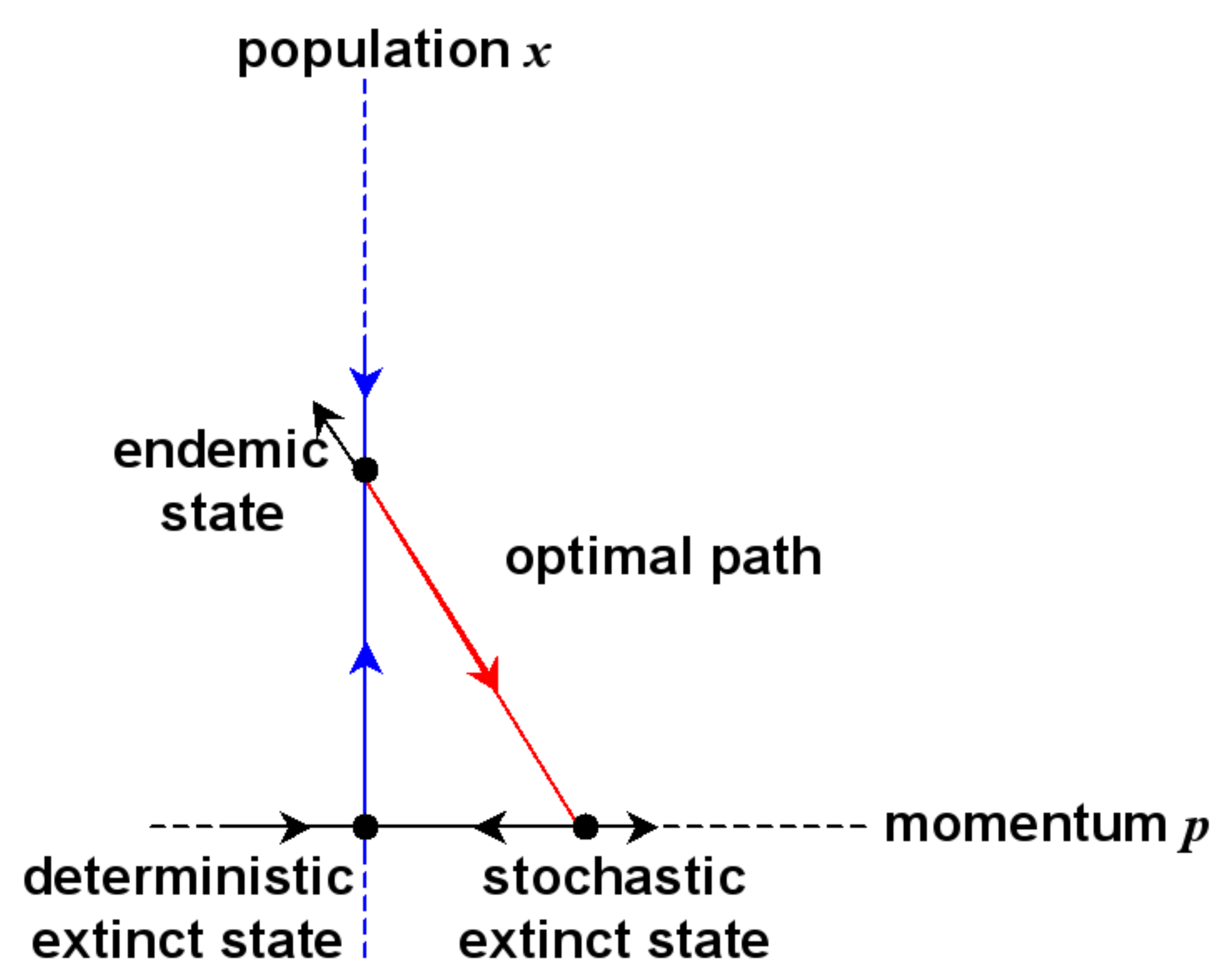}
 \includegraphics[width=6cm]{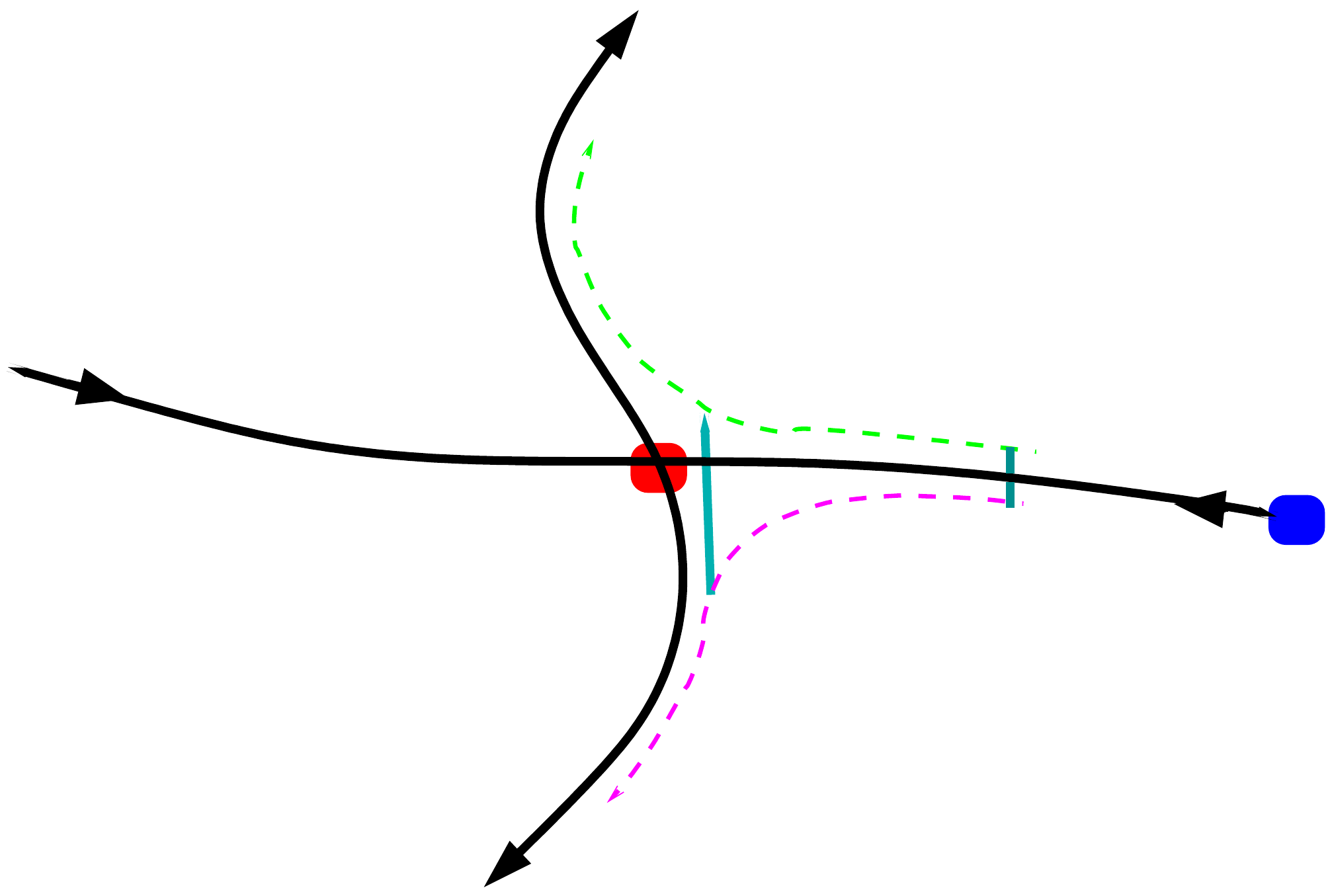}
 \caption{a) Schematic showing the fixed points and heteroclinic
   trajectories (trajectories connecting fixed points).  Coordinate axes are dashed lines. The noise coordinate
     is indicated by the momentum ($p$) coordinate. The deterministic manifold ($p=0$) is indicated in blue.  Deterministically, the extinct state is repelling and endemic state is attracting. However, the endemic state has unstable directions for nonzero noise ($p \neq 0$), and the optimal path (red) is the trajectory carrying the system from the endemic state to a stochastic extinct state.  b) Schematic showing the path
     from the endemic state (blue) to the extinct state (red).
 The optimal path leaves the
     endemic point along an unstable manifold, and connects to the extinct state
     along its stable manifold. The magenta and green dashed lines represent
     trajectories initially separated by the optimal path.  The initial starting
     distance between trajectories near the endemic state expands exponentially
     in forward time (shown by the cyan lines).  Locally, this shows that the finite-time Lyapunov measure of
     sensitivity with respect to initial data is maximal along the optimal
     path.}\label{fig:hyperbolic}
 \end{center}
 \end{figure}

In yet another application, knowledge of the optimal path to extinction
has the advantage that, in the presence of noise (that is estimated
from data) and a known population of infectious individuals, it may be possible to develop better
vaccine controls that  reduce the time to extinction. Figure~\ref{fig:hyperbolic}(a)
shows a schematic of the path to extinction, where the extinct state
is a saddle with stable and unstable directions. For many
epidemic models, the extinct state can be shown to have a similar
geometry of stable and unstable directions. An approach to the extinct state on the optimal path will lead to the
fastest time to extinction. Moreover, since the extinct state has a saddle structure
in the presence of noise, it may be controlled with projection methods~\citep{ST1994,SCT1997} or probabilistic techniques~\citep{Schwartz2004}.

One can consider instituting a method of parameter control, where the
parameters could be vaccine levels or treatment of infectious individuals.
Combined with the above-mentioned monitoring techniques, the control method
will allow one to move an existing state that deviates from the optimal
path closer to the optimal path as time
evolves.  By adjusting the parameters, we may target the stable
directions of the path when we are close to epidemic extinction~\citep{ST1994}. Comparing
observations with model predictions of the optimal path allow us to
use controls to minimize the time to extinction.

\begin{acknowledgements}
The authors gratefully acknowledge support from the Office of Naval
Research, the Air Force Office of Scientific Research, the Army
Research Office, and the
National Institutes of Health.  I.B.S and L.B.S. are supported by Award
Number R01GM090204 from the National Institute Of General Medical
Sciences.  The content is solely the responsibility of the authors and
does not necessarily represent the official views of the National
Institute Of General Medical Sciences or the National Institutes of Health.
We also gratefully acknowledge M.~Dykman for helpful discussions.
\end{acknowledgements}

\appendix{Theory of large fluctuations}\label{sec:app-tlf}
Letting $N$ denote the population size, the state variables
$\bm{X}\in \mathbb{R}^{n}$ of the system describe the components
of a population, while the random state transitions which govern the
dynamics are described by the transition rates $W(\bm{X},\bm{r})$,
where $\bm{r}\in \mathbb{R}^{n}$ is an increment in the change of
$\bm{X}$. In the large population limit without any fluctuations, the
mean field equations are given by the system $\dot{\bm{X}}=\sum_{\bm{r}}\bm{r} W(\bm{X},\bm{r}).$

Consideration of the net change in increments of the state of
the system results in the following master equation for the probability
density $\rho(\bm{X},t)$ of finding the system in state $\bm{X}$ at time $t$:
\begin{equation}
\frac{\partial\rho(\bm{X},t)}{\partial
  t}=\sum_{\bm{r}}\left[W(\bm{X}-\bm{r};\bm{r})\rho(\bm{X}-\bm{r},t)-W(\bm{X};\bm{r})\rho(\bm{X},t)\right].\label{e:MasterEquation}\\
\end{equation}

The solution to the master equation~\eqref{e:MasterEquation} has
  a peak around a stable steady state in the limit of large population ($N \gg
1$) with width $\propto N^{1/2}$~\citep{dyscla08,sbdl09}. However, since we
are interested in the probability of extinction, we will consider the tail of
the distribution, which gives the probability of having a small number of
individuals. The tail of the distribution can be obtained by employing
the ansatz given by
equation~\eqref{e:Action}, which is an assumption for how the probability density scales with population size~\citep{Kubo1973,dyk90,dyscla08}.
Equation~\eqref{e:Action} also implies that a maximum in
the extinction probability can be found minimizing the action over a
set of extinction paths starting from the stable endemic state.  The
  assumption of this functional form for $\rho$
 allows the action to be derived from properties of the master equation.

The density, $\rho(\bm{X},t)$, can be found by substituting the ansatz given by equation~\eqref{e:Action} into the master equation.
The resulting equations for the action are given by the
Hamilton-Jacobi equation for a Hamiltonian, $H$,   given by
$\partial{\cal{S}} / \partial t + H({\bm x},{\partial{\cal{S}}/ \partial{\bm
    x} };t) = 0,$ where we have normalized the population components  and
rates respectively as $\bm{x}=\bm{X}/N, w(\bm{x},\bm{r})=W(\bm{X},\bm{r})/N$.

Following the classical mechanics convention, define a conjugate momentum to
the state space, $\bm{x}$, by letting  ${\bm p} =\partial{\cal{S}}/
\partial{\bm x}$ and  where $H({\bm x},{\bm p};t)$ is the classical
Hamiltonian~\citep{Kubo1973}.  The Hamiltonian, $H$, depends both on the state of
the system, ${\bm x}$, as well as the momentum, ${\bm p}$, which provides an
effective force due to stochastic fluctuations on the system.  The Hamiltonian
equations of motion provide trajectories in time of ${\bm x}(t)$ and ${\bm
p}(t)$, and as such, describe a set of paths going from an initial
point at time $t_i$ to some final point at time $t_f$ in $(\bm{x},\bm{p})$
space. For a given path,  the action
is given by ${\cal S = }\int_{t_{i}}^{t_{f}}\bm{p}(t)\dot{\bm{x}}(t)dt$,
and as such determines the exponent of the  probability of observing that path.    (It should be noted
that instead of using the Hamiltonian representation, one could
use the Lagrangian representation
$L(\bm{x},\dot{\bm{x}};t)=-H(\bm{x},\bm{p};t)+\dot{{\bm x}}\cdot{\bm p}$,
which results in an equivalent solution.)

The action reveals much information about the probability evolution of the system, from scaling
near bifurcation points in non-Gaussian processes to rates of extinction as a
function of epidemiological parameters~\citep{dyk90,dyscla08}.  As
 already stated, in order to maximize the
probability of extinction, one must minimize the action $\cal{S}$.
The minimizing formulation entails finding the solution to the
Hamilton-Jacobi equation, which means that
one must solve the $2n$-dimensional system of Hamilton's equations [$\dot{\bm
  x}=\partial_{\bm p}H({\bm x},{\bm p})$, $\dot{\bm p}=-\partial_{\bm
  x}H({\bm x},{\bm p})$] for $\bm{x}$ and
$\bm{p}$, where the Hamiltonian is explicitly given as
\begin{equation}
H(\bm{x},\bm{p},t)=\sum_{\bm{r}}w(\bm{x},\bm{r})[\exp{(\bm{p}\bm{r})}-1].\label{e:hamilonian}
\end{equation}
The appropriate boundary conditions of the system are such that a
solution starts at a non-zero state, such as an endemic
state, and asymptotically approaches one or more zero components of
the state vector, representing a disease free state.  Therefore, a trajectory that
is a solution to the two-point boundary value problem determines a path, which in
turn yields the probability of going from the initial state to the final state. The optimal
path to extinction is the path which minimizes the action
in either the Hamiltonian or Lagrangian representation.

We compute the trajectory satisfying the Hamiltonian system that has as its asymptotic
limits in time the endemic state as $t\to-\infty$ and the extinct
state as $t\to+\infty$.  The momentum $\bm p$
represents the force of the fluctuations on the population, and this momentum
changes the stability of the equilibrium points.  Both the endemic and extinct states have
attracting and repelling directions for ${\bm p} \ne 0$, as shown schematically in Figure \ref{fig:hyperbolic}.

Given optimal path trajectories $(\bm{x}(t),\bm{p}(t))$, the action
with correct limits is found by
${\cal S}(\bm{x})=\int_{-\infty}^{\infty}\bm{p}\bm{\dot{x}}\,
dt$~\citep{Goldstein2001}.


 \clearpage

 \end{document}